# Potential utilization of Battery Energy Storage Systems (BESS) in the major European electricity markets

Yu Hu [1]*, Miguel Armada [2], María Jesús Sánchez [2]

[1] *Simulyde S.L., Madrid, Spain.*

[2] *Escuela Técnica Superior Ingenieros Industriales, Universidad Politecnica de Madrid, Madrid, Spain.*

## Abstract

Given the declining cost of battery technology in the last decade, nowadays BESS becomes a more attractive solution in electrical power systems. The objective of this work is to analyze the potential utilization of BESS in the major European electricity markets. A general payoff model for BESS operation is proposed to correctly address the operational flexibility of battery systems. Utilization factors such as potentially profitable utilization time and rate are calculated for common applications including energy arbitrage and frequency support services using real market information. The result shows that under the current empirical estimation of the battery cost and lifetime, BESS is not feasible for energy arbitrage in most of the European electricity markets. However, BESS shows clearly and significantly higher potential in providing frequency support services. The result suggests that, when the frequency containment reserve is remunerable, the potentially profitable utilization of BESS has become already accretive in most of the European countries. For example from January to September 2021, the potentially profitable utilization rate has reached almost 100% for the FCR-N service in the Danish market. Comparing the regional electricity markets in Europe, BESS has shown significant potential in becoming a feasible solution in Central Western Europe and parts of Northern Europe by providing frequency regulation services. Meanwhile, in the British Isles and some other islanded local markets, a remarkable level of scarcity of flexibility has been revealed by the investigation, and the potential of BESS would also be considerably encouraging.

## Keywords

---

[1] Corresponding author. Tel: +34-917395257; Email: yu.hu@invesyde.com

# Highlights

- A general payoff model is proposed to address the operational flexibility of battery systems
- Potential utilization of battery systems are evaluated using utilization factors
- Battery systems are not feasible for energy arbitrage in the current European electricity markets
- Potential utilization of battery systems is promising in Europe for frequency regulation services

# 1. Introduction

Nowadays, intense concern about climate change is increasing among policy-makers and other stakeholders in many major European economies. In July 2021, the European Commission unveiled a set of legislative proposals aimed at achieving carbon neutrality by 2050, while reducing emissions by 55% by 2030 from the 1990 level as an intermediate target [1]. And in the United Kingdom (UK), the government published, in November 2020, a 10 Point Plan to support the economy and employment market during the transition process to net-zero emission by 2050 [2].

As a replacement of fossil fuels fired power plants, expanding and integrating renewable energy capacity is one of the most essential actions in these ambitious plans. In the new proposal by the European Commission [3], the binding target for the share of renewables increased to 40% by 2030, 8% higher than the existing target of 32% established in the 2018 directive. And in the UK, the target for offshore wind power capacity is set at 40GW by 2030, three times higher than the current installed capacity in 2020 [2].

The flexibility issues would arise and become crucial with higher penetration of intermittent renewable energy sources in the energy mix. Given the fact that a great share of dispatchable generation capacity based on fossil fuels would be replaced by renewable energy, energy storage, as an alternative flexibility provider, is considered as a critical resource to achieve the sustainability goals for the future energy system [4]. Among all the energy storage technologies, battery technologies, especially the Li-ion battery, have experienced considerable cost reduction in the last years. Nowadays, Battery Energy Storage Systems (BESS) becomes more attractive in providing flexibility with decentralized and distributed solutions. According to Hu et al. [5], although BESS might not be completely economically feasible in some deregulated electricity markets such as the Spanish one, the potential utilization for BESS is clearly higher in countries where higher flexibility is demanded. At the same time, the potential application of BESS would be more promising in ancillary services than energy-based applications such as energy arbitrage.

The objective of this work is to review the current market applications and analyze the potential utilization of BESS in the major European electricity markets. In order to evaluate the potential utilization, a general model for BESS operation in electricity markets is proposed to correctly address the operational flexibility of battery systems. And utilization analysis is applied in the major European electricity markets using the concept "potentially profitable utilization time" [5]. Since the battery installation cost compared with the limited lifetime of the battery itself remains as the main barrier for feasible BESS applications, the potential changes in BESS utilization emerging from the battery cost reduction are also analyzed, which provides essential information for future scenarios.

This paper contributes to the knowledge of the BESS operation under real deregulated electricity markets with the proposed general BESS payoff model. At the same time, by addressing the potential decline of battery cost as future scenarios, and comparing the result across different regional markets with different applications, this study provides a general overview of future potential utilization for battery systems in the major regional European electricity markets. The results of this study also contribute to the awareness of flexibility provided by BESS in electricity markets, which would provide essential information for market participants such as system/market operators, policy-makers, investors, and project developers.

The remainder of this paper is organized as follows. Section 2 provides a general review related to the current electricity market design in European countries. Section 3 summarizes the current situation for BESS in Europe and common applications for BESS as a flexibility provider in the state of the art. Section 4 presents the proposed BESS operation model and the potential utilization analysis of BESS in the major European electricity markets. Finally, Section 5 discusses the related issues and presents the main conclusions of this paper.

## 2. Electricity markets in European countries

### 2.1. Market structure and market integration

The electricity market liberalization in Europe was initiated in 1996 with the enactment of the first European Directive concerning the liberalization of the electricity market [6]. During the market liberalization process of its member countries, the European Union (EU) has also adopted several regulations to achieve a single internal electricity market ensuring secure and affordable energy supplies [7]. According to Bunn and Gianfreda [8], the level of integration across the different regional and national electricity markets has been significantly improved in the last decades.

Currently, according to the European Network of Transmission System Operators (ENTSO-E), the electricity market in European countries is structured into several individual markets: the forward market, the day-ahead market, the intraday market, the ancillary services market, and the imbalance settlement [9].

In the forward market, market participants trade mid-long term products or electricity derivatives to hedge the risks and uncertainties in the day-ahead market. According to the European Union Agency for the Cooperation of Energy Regulators (ACER) and Council of European Energy Regulators (CEER), the liquidity of the forward markets in European countries diverges significantly: bidding zones such as FR (France), DE-AT-LU (Germany/Austria/Luxembourg, and the bidding zone has become DE-LU since October 2018), and GB (Great Britain) are with the highest level of liquidity, however, the forward market liquidity in other large bidding zones such as Spain and Poland are much lower [10].

The day-ahead market is the major market where power delivery commitments are settled between power producers and consumers for the next 24 hours. The market clear prices in this market are normally considered as the most important reference for energy prices. In the EU, the Single Day-ahead Coupling (SDAC) initiative has been established to create a single pan European cross zonal day-ahead electricity market [11]. Currently, the PCR EUPHEMIA algorithm (acronym of Price Coupling of Regions and Pan-European Hybrid Electricity Market Integration Algorithm) has been commonly adopted in Europe to calculate electricity prices and to allocate cross-border transmission capacity [12].

When the day-ahead market is closed, market participants may make short-term adjustments in the intraday markets. In general, intraday markets are organized as collective auctions or based on continuous trading. Several European countries have initialized the Cross-Border Intraday (XBID) project to integrate the national intraday markets. Currently, 14 countries have participated in this project, including, Austria, Belgium, Denmark, Estonia, Finland, France, Germany, Latvia, Lithuania, Norway, the Netherlands, Portugal, Spain, and Sweden.

In the ancillary services market, also referred to as the balancing market, service providers such as power generators, demand response facilities, and BESS may offer ancillary services to the system operator and be remunerated if the offer is accepted and the services are correctly provided. Meanwhile, during the imbalance settlement process, balance responsible parties such as electricity producers, consumers are financially responsible for the imbalances of their portfolios [9].

Ancillary services are a set of services related to the security and reliability of a power system [13], which include frequency control, voltage support, black start services, automatic islanding, etc [14]. In this work, we focus on frequency control services since a high level of flexibility in active power (or real power) is required in these services, which are ideal for BESS installations.

In order to maintain a stable system frequency, the ancillary services for frequency control in Europe are generally organized in the following levels: The Frequency Containment Reserve (FCR), the automatic Frequency Restoration Reserve (aFRR), manual Frequency Restoration Reserve (mFRR), and Replacement Reserves (RR).

The FCR, also commonly known as primary control [15], aims at "the operational reliability of the power system of the synchronous area and stabilizes the system frequency at a stationary value after a disturbance or incident in the time-frame of seconds, but without restoring the system frequency and the power exchanges to their reference values." The common activation time for the FCR is typically within 30 seconds. The EU has established a project named "FCR Cooperation" to integrate the market of FCR procurement. And until October 2021 eight countries (Austria, Belgium, Slovenia, Germany, Switzerland, Western Denmark, France and the Netherlands) have joined the project.

The aFRR (also known as the secondary reserve) and the mFRR (also known as the tertiary reserve) will be activated if the frequency deviation lasts. For example, according to

ENTSO-E [9], the automatic activation time for aFRR is between 30 seconds to 15 minutes, and for the mFRR, the activation could be semi-automatic or manually controlled and the time limitation is within 15 minutes. In other words, if the frequency deviation lasts for more than 30 seconds, the aFRR service is activated to replace the FCR. The same, for frequency deviations which are more than 15 minutes, the mFRR is activated to replace the aFRR.

The service of RR exists only in some of the European countries. Similar to the aFRR and mFRR, the RR is designed for the replacement of the activated aFRR and/or mFRR services, and the required activation time is more than 15 minutes [9].

Platforms such as PICASSO (Platform for the International Coordination of Automated Frequency Restoration and Stable System Operation), MARI (Manually Activated Reserves Initiative), and TERRE (Trans-European Replacement Reserves Exchange) have been implemented to create an integrated platform for the aFRR, the mFRR, and the RR respectively. Nevertheless, at the moment the market design for the services mentioned above in each country remains dissimilar and uncoordinated.

## 2.1 Regional electricity markets

Considering the geographical distribution and the interconnection capacities, the European electricity market is mainly composed of several interconnected regional markets [16], including Central Western Europe (Austria, Belgium, France, Germany, Luxembourg, the Netherlands, Switzerland), the British Isles (Great Britain, Ireland), Northern Europe (Denmark, Estonia, Finland, Latvia, Lithuania, Sweden, Norway), the Apennine Peninsula (Italy, Malta), the Iberian Peninsula (Spain and Portugal), Central Eastern Europe (Czechia, Hungary, Poland, Romania, Slovakia, Slovenia), and South Eastern Europe (Bulgaria, Croatia, Greece, and Serbia). Compared with the electricity markets in Eastern Europe, the Central Western Europe, the British Isles, Northern Europe, the Apennine Peninsula, and the Iberian Peninsula are further integrated not only by the day-ahead market but also by the intraday market and the balancing market [17].

In this work, the national electricity markets of Denmark (Northern Europe), France (Central Western Europe), Germany (Central Western Europe), Great Britain (British Isles), Italy (Apennine Peninsula), Norway (Northern Europe), and Spain (Iberian Peninsula) are selected and selected as the "major European electricity markets", due to the economic impact of these countries in Europe and the representativeness and market share in their respective regional market.

## 3. BESS in Europe: current situation and market applications

According to the recently published report from the European Commission [18], although the dominating energy storage reservoir in Europe is still pumped hydro storage, new batteries projects are being developed rapidly especially in Germany and the UK. The report states that the Lithium-ion batteries represent most of BESS projects. Figure 1 presents the current installed capacity for BESS systems in different European countries. Meanwhile, in terms of future projection, the UK presents the most important power capacity, followed by Ireland and Germany. Up to the publication of the report, the total power capacity of authorized projects with electrochemical storage is 5499 MW in the UK, almost 10 times higher than the current operating capacity.

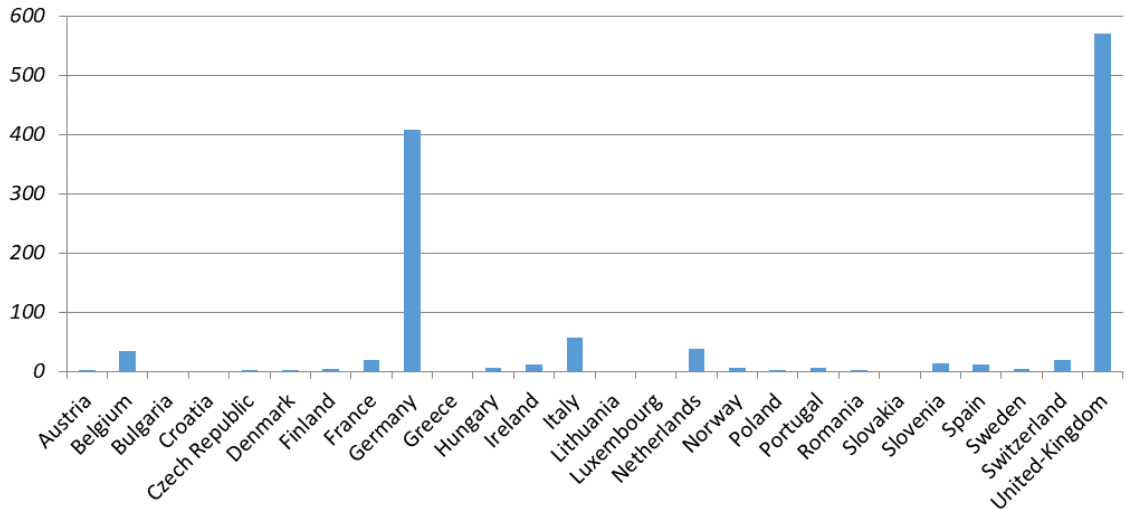

Figure 1. Total installed power capacity (in MW) of operated BESS in European countries [18].

Flexibility is considered essential for the electrical power system in the future with a much higher penetration level of renewable energy. Although the potential applications of BESS, as a distributed and decentralized technology, has been investigated by many research works. In this study, we focus only on the applications based on the existing European electricity market or ancillary services that are currently demanded by the system operator. Among all the market applications, energy arbitrage [19][20] and frequency regulation [21] are the most common form of BESS application in electricity markets. Energy arbitrage refers to the benefit obtained from the price difference of the whole market: charging during low-price periods, discharging during high-price periods. Frequency regulation, on the other hand, requires BESS to deliver dynamic power in response to frequency deviation of the system or in response to automatic or manually controlled reserve services [22]. Apart from the two main market applications mentioned above, other BESS solutions in the power grid may include but not limited to congestion management [23], end-user applications [24][25], isolated power grid [26], local markets [27], peak-load shaving [28], and hybrid system [29].

Focusing on energy arbitrage application, Núñez et al. [30] calculated and compared the Net Present Value (NPV) and Internal Rate of Return (IRR) for different European countries in 2019 and concluded that the application is not economically feasible at that time. Among all the countries with negative economic indicators, the most promising countries would be the United Kingdom and Ireland, while Spain and Portugal showed the worst performances. The result aligns with other previous research works [31][32], which generally argue that arbitrage in itself cannot support investments in the energy storage sector.

And for frequency regulating services and mixed applications, in the UK, Gundogdu et al. [22] developed a control algorithm for maximizing the profit in Frequency Response Services and Energy Arbitrage in the UK electricity markets, the control parameters have been validated using existing BESS platforms.

Thien et al. [33] investigated operational requirements of the German FCR market and simulated the BESS operation with a 5 MW/ 5 MWh battery storage system, at the same time operational strategies are proposed and validated with the system frequency of the year 2014. Münderlein et al. [34] introduced different marketing strategies providing frequency control service in Germany using the same BESS platform, and the result shows considerable profitability in the FCR market. Koller et al [35] tested different applications including the primary frequency control using a 1MW/580MWh BESS in Zurich, Switzerland. Oudalov et al. [36] developed an optimization algorithm for primary frequency control using BESS.

In Nordic electricity markets, Heine et al. [37] and Hellman et al. [38] tested an industrial-sized BESS for providing frequency control services in Finland and proved that the availability of the battery system could be considerably improved with optimized charging and discharging strategies. Also focusing on the Finish electricity market Divshali and Evens [39] studied the bidding strategy of BESS in the frequency containment reserve (FCR) market with empirical data. Filippa et al. [40] simulated the economic performance of BESS providing Frequency Containment Reserve for Normal operation (FCR-N) under Eastern Denmark's (DK2) regulations. Ahčin et al. [41] performed a techno-economic analysis of BESS in the Norwegian electricity market and concluded that compared with other services, frequency regulation is currently the most feasible application.

The main limitations in the currently available literature are listed as follows. Firstly, most related research works are conducted for specific applications/services with specific market conditions, and the operational strategies of BESS in these works are often highly optimized accordingly. However, it becomes difficult to apply those proposed models to other similar applications or markets. Thus, a general across-the-board model for BESS operation is extremely necessary for common market applications in deregulated electricity markets.

Secondly, deterministic valuation modeled such as "net present value" or "payback periods" are normally calculated in the current feasibility studies and economic analyses. However, due to the operational flexibility of the BESS, a battery would only be operated when the market price spread or the spread between the service remuneration and the corresponding cost is positive. As a result, similar to the dark/spark spread options for thermal power plants, option-based models would be more appropriate for BESS operation modeling.

Thirdly, since most of the research works are focusing on specific applications, the result obtained would not be easily compared among different markets or even among different applications. Meanwhile, the current studies are not able to provide a general picture of the future potential utilization of BESS for grid applications.

In order to shape a future pathway of BESS in Europe and correctly address the operation flexibility, a general option-based BESS operation model is presented with utilization analysis conducted for the major European electricity markets. Utilization analysis is a common method for measuring the potential for energy projects. Concepts such as "Equipment Utilization Hours [42]" and "Degree of Utilization [43]" are widely used for feasibility studies in wind, solar, and hydropower projects. In this paper, rather than measuring by total energy delivered from the device, the utilization factor "potentially profitable utilization time" proposed by Hu et al. [5] has been applied in this paper.

The advantage of the proposed method would include, on the one hand, the proposed utilization factors would be more effective in measuring the future market potential for different stakeholders such as system operators, policy-makers, and technology promoters even though the project is not economically feasible. On the other hand, the potentially profitable utilization time is a valuable measurement to compare the market potential for different applications with different remuneration structures [5].

## 4. Potential utilization of BESS in the EU

### 4.1 General BESS payoff and operation model in electricity markets

We present a general BESS operation model taking into consideration the operational flexibility of the system. In deregulated electricity markets, BESS gets paid by selling the stored energy or by providing services of flexibility. On the contrary, the expenses shall include the battery wear cost and the operational costs. Meanwhile, the state of charge of BESS needs to be balanced after providing a certain service or application. Due to the operational flexibility, BESS assets are only operated when it is profitable. Thus the payoff for a certain time period t for a specific application can be written as

$$\text{Payoff}_t = \max(S + B - O, 0),$$

where,
    S is the remuneration of the service,
    B is the balancing term,
    O is the operating expenses, including the battery wear cost and other expenses.

Note that this formula above could be considered as an option liked model where the remuneration of the service and the balancing term contain stochastic variables such as energy prices, while the operational expenses could be taken as the strike price. The "storage option" would only be excised when the payoff value is higher than zero.

In fact, the payoff of BESS applications depends on the specific market design. The service remuneration may include different terms, in general, the service payments in Europe are normally based on power capacity, in national currency per MW, and/or energy supported, in national currency per MWh. At the same time, a BESS may receive positive remuneration through the energy balancing in the case that the BESS is discharged in the balancing phase. In conclusion, the service rumination term S and the balancing term B now become,

$$S = P_c C_s + P_{sd} E_{sd} - P_{sc} E_{sc},$$

$$B = P_{bd} E_{bd} - P_{bc} E_{bc},$$

where,
- $P_c$ is the capacity price for battery service,
- $C_s$ is the power capacity provided,
- $P_{sd}$ is the discharge price for battery service,
- $P_{sc}$ is the charge price for battery service,
- $E_{sd}$ is the discharged energy for battery service,
- $E_{sc}$ is the charge energy for battery service,
- $P_{bd}$ is the discharge price for energy balancing,
- $P_{bc}$ is the charge price for energy balancing,
- $E_{bd}$ is the discharged energy for energy balancing,
- $E_{bc}$ is the charge energy for energy balancing,
- $\eta$ is the round-trip efficiency.

For BESS assets, a round-trip efficiency is defined as the energy required being charged beforehand, in kWh for example, to generate one kWh of electricity by discharging the battery. The round-trip efficiency is one of the key characters of specific BESS: the higher the ratio, the more efficient the battery system and the less the energy loss during the storage process. If we consider that the battery status is balanced for the service, there is,

$$E_{sd} + E_{bd} = E_{sc} \cdot \eta + E_{bc} \cdot \eta.$$

On the other hand, the lifetime of batteries is not infinite due to the degradation caused by every charge-discharge cycle [44]. Also, the battery degradation may also increase the internal resistance and decrease the round-trip efficiency [45]. Once the chemicals inside a battery cell are exhausted, the cell would need to be replaced. For modeling purposes, the lifetime of battery systems is commonly described as cycle life, which is defined as the number of complete charge-discharge cycles that the battery can perform before it gets replaced or its nominal capacity falls below a certain level of its initial rated capacity [46]. Accordingly, the cost of the delivered energy from the battery can be calculated as the battery cost per kWh divided by the Equivalent Full Cycles until the battery is replaced [47]. Therefore, the term of battery wear cost O can be calculated as,

$$O = W_d(E_{sd} + E_{bd}) + \text{opex},$$

where,
    $W_d$ is the battery wear cost per unit of energy delivered from the battery,
    opex is the term for other marginal operating expenses.

In order to estimate the battery wear cost, the following references are taken into consideration. For the common Li-ion batteries, Zubi et al. estimated the battery specific cost to be around 300 €/kWh with a cycle life around 2000 cycles [48]. And for real battery storage products, Tesla provides a warranty that guarantees 70% energy retention in 10 years with a maximum number of 2800 Equivalent Full Cycles for its Li-ion battery product "Powerwall" [49]. If we consider an empirical battery storage cost of 300 €/kWh and a conservative estimation of 3000 equivalent full cycles lifetime before the battery is replaced, it would imply a pure battery wear cost of 0.1 €/kWh (100 €/MWh). At the same time, since it is commonly expected that the battery cost would decline with an increased battery lifetime in the following years, in this study, we consider the battery wear cost ranges from 0 to 0.1 €/kWh (0 to 100 €/MWh).

In this work, a utilization analysis is conducted by comparing the potentially profitable utilization time of BESS for grid applications in different European countries. The Potentially Profitable Utilization Time (PPUT) defined by Hu et al. [5], can be calculated as the total time periods that the payoff is higher than zero, where,

$$\text{PPUT} = \sum_t(\text{Payoff}_t > 0).$$

Accordingly, the Potentially Profitable Utilization Rate (PPUR) of BESS for a specific time period can be defined as the potentially profitable utilization time during this time period divided by the whole time period,

$$\text{PPUR} = \frac{\sum_t(\text{Payoff}_t > 0)}{T},$$

where T is the total time period.

### 4.2 Energy arbitrage and day-ahead markets

One of the most discussed applications for BESS is the energy arbitrage, where BESS buys energy when the electricity price is low and sells electricity during peak hours when the price is high. If the price signal is correct, BESS is supposed to store electricity when renewable energy production is relatively high and the power load is relatively low, and vice versa.

In this work, the operational frequency energy arbitrage is considered as daily cycles. In other words, the battery system will perform one full charge-discharge cycle daily. The reasons have been discussed by Hu et al. [5] Firstly, due to the demand and solar pattern, a daily pattern of electricity price with peak and valley hours is generally observed. Secondly, the daily operation cycle aligns with the designed operation pattern in terms of the designed lifetime and warranty provided by the manufacturer. And lastly, with common BESS designs which store energy for only a few hours, it would not be meaningful to perform energy arbitrage in longer time periods.

For energy arbitrage the remuneration of the application is solely the energy discharged during the peak hours, the BESS operation and payoff function can be written as follow (for the sake of simplicity, it is assumed that the other operational cost has been included as the battery wear cost),

$$\text{Payoff}_d = \max(P_{sd}E_{sd} - P_{bc}E_{bc} - W_d E_{sd}, 0),$$

$$E_{sd} = E_{bc} \cdot \eta,$$

where,
  $\text{Payoff}_d$ is the daily payoff of BESS for energy arbitrage,
  $P_{sd}$ and $P_{bc}$ are the hourly prices when BESS chooses to perform energy arbitrage.

In this work, historical market prices (Jan 2019 – Sep 2021) from the day-ahead market (obtained from ENTSO-E) have been used to conduct the analysis. Since BESS would attempt to maximize the profit, the daily maximum and minimum hourly prices are selected as the prices to discharge and charge the storage asset respectively. In other words, it is assumed that the BESS is with perfect price information. Despite the fact that it is not practically achievable for the BESS operator to perfectly predict the price in the day-ahead market, short-term price models can be used in revealing useful information about the price difference, and the energy positions could be corrected in the intraday market.

Figure 2 presents the potentially profitable utilization rate for energy arbitrage under different battery wear costs from January 2019 to September 2021 in the countries being studied, in Days. It should be noted that in some of the countries, there are several bidding zones with different market prices. In Figure 2, bidding zones "IT-Centre-South", "DK1", and "NO1" are selected for Italy, Denmark and Norway. Detailed analysis focusing on the effect of bidding zones will be presented after. And for the market in the UK, the bidding zone GB (Great Britain) is selected, and a conversion rate of 1 GBP = 1.12 EUR has been considered.

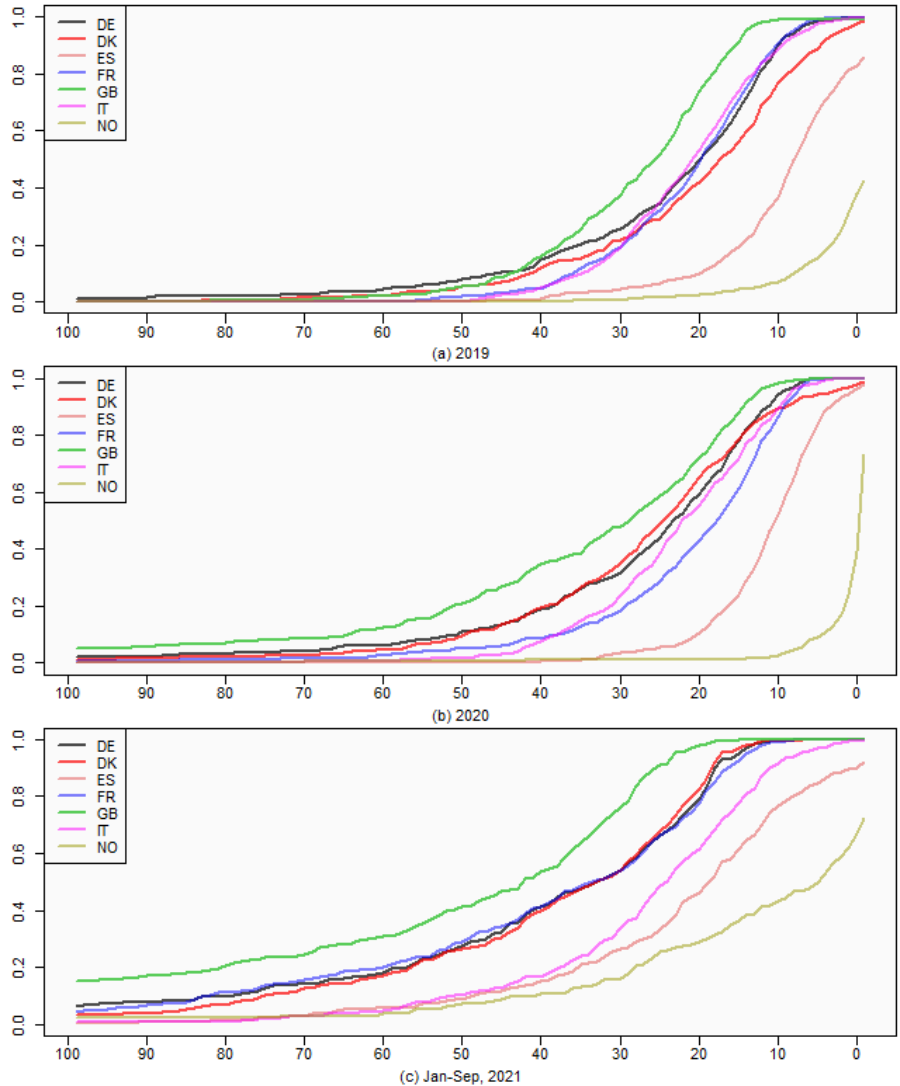

**Figure 2. Potentially profitable utilization rate (Y-Axis) for energy arbitrage under different battery wear cost (in €/MWh, X-axis) in the major European electricity markets.**

The interpretation of curves would represent the potential utilization of BESS given the conditional cost reduction in the future. When a curve is more to the right side, it means that the battery cost required is lower in order to obtain benefits. In contrast, a curve shifted to the left indicates that, even with higher battery costs, it is possible to profitably operate BESS for a certain number of days.

It can be observed that the battery utilization cost required for a BESS to be potentially profitable varies greatly among the European countries. Under a current empirical battery wear cost (approximately 100€/MWh), energy arbitrage will not be profitable for most of the countries in the EU. However, if the battery wear cost and other variable cost decreases below 40-50€/MWh, the potentially profitable utilization time will start to increase rapidly for most of the European countries and local markets such as Great Britain, France, Germany, and Italy. Other countries such as Spain would require a much lower battery cost, lower than 20€/MWh, for energy arbitrage. Lastly, for Norway, energy arbitrage is hardly profitable even with a very low battery wear cost.

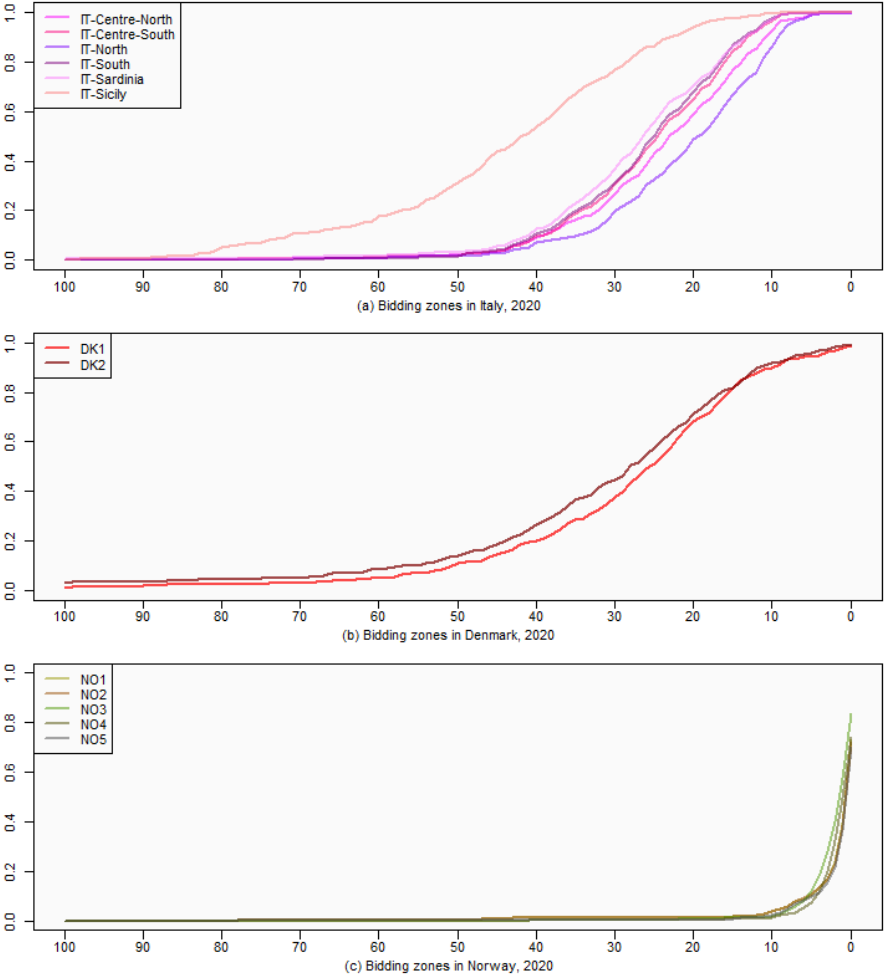

**Figure 3 Potentially profitable utilization rate (Y-Axis) for energy arbitrage under different battery wear cost (in €/MWh, X-axis) in Italy (a), Demark (b), and Norway (c), in 2020.**

As mentioned above, in some countries there are several separate bidding zones, and the remuneration of BESS would be different depending on the location. Figure 3 presents the potentially profitable utilization time for energy arbitrage under different battery wear costs in 2020 in Italy, Denmark, and Norway. In most cases, the curves of potentially profitable utilization time are similar in the same country. However, the bidding zone Sicily in Italy shows much higher potential utilization for BESS in energy arbitrage applications.

The high potential utilization for BESS could be also overserved in other regional markets with limited interconnection capacity. Figure 4 presents the potentially profitable utilization time for energy arbitrage under different battery wear costs in Great Britain and Ireland. In Ireland, the potential utilization of BESS in energy arbitrage would be even more profitable than in Great Britain for given battery wear costs.

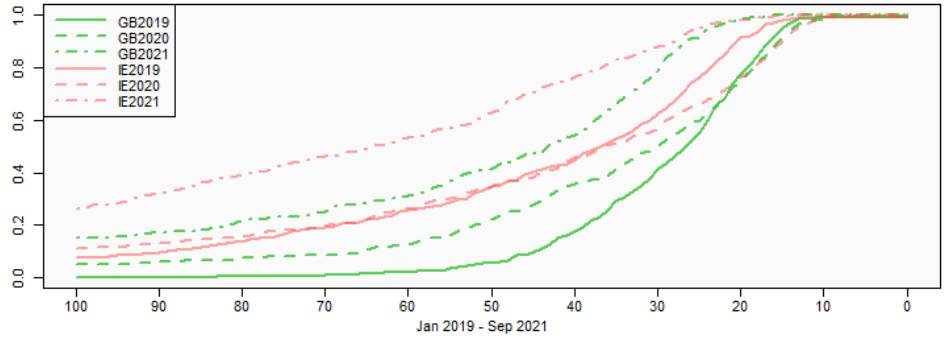

**Figure 4. Potentially profitable utilization rate (Y-Axis) for energy arbitrage under different battery wear cost (in €/MWh, X-axis) in Great Britain and Ireland, Jan 2019 to Sep 2021**

The result could be explained by the following two factors: the flexibility in the generation structure and the interconnection with the neighboring countries. On the one hand, countries such as Spain and Norway present high flexibility due to their generation structure, in which hydraulic power generation represents a significant percentage of total generation. However, countries such as France and Germany are highly dependent on the power generation technologies which are less flexible, such as nuclear power and coal power.

On the other hand, the power interconnection also plays a key role that affects the price fluctuation in the day-ahead market. The price signal of the larger market player may have a significant impact on the regional market. And local markets get merged if the interconnection capacity among countries is sufficient enough, for example, Germany and Luxemburg are sharing the same bidding zone. This would explain the similar utilization curves among the central European countries. On the contrary, countries such as Great Britain and Ireland, despite having certain interconnections with other European countries, are relatively isolated from continental Europe. Higher market price volatility has been observed in these relatively isolated local markets since the flexibility demand cannot be supplied by alternative flexibility resources in the neighboring markets.

### 4.3 Frequency containment reserve

Unlike the day-ahead market, up to the present, there is no common market design for the frequency support services across the European countries. For the FCR service (also known as the primary reserve), the remuneration structure varies in terms of capacity payments, energy payments and market clearing methods. Meanwhile, the technical requirements also vary in different terms such as frequency band and response time requirements. Table 1 presents a summary of the FCR services for the countries being studied. Note that in some countries, the FCR market could be different depending on the bidding zone and supported by several separated products.

**(Insert Table 1 about here)**

At a regional electricity market level, the primary reserve services (FCR) in the Apennine Peninsula and the Iberian Peninsula (e.g, Spain, Portugal and Italy) are mandatory and not remunerable. And in Central Western Europe (Denmark-DK1, France and Germany), thanks to the FCR Cooperation project, the FCR markets have been more integrated recently. Also, similar services and products are shared for the FCR markets in Northern Europe (Denmark-DK2 and Norway). Lastly, in the British Isles, the FCR service is complicated since the market integration level is relatively low. For example, in the UK the "Firm Frequency Response" is tender based and the procuring of the service is based on case-by-case studies. Also, the UK has launched another service named "Enhanced Frequency Response" focusing on energy storage facilities in 2016. However, the service is contracted for only 4 years and there would be no future auctions.

In this work, the potential utilization analysis is conducted for the remunerable and standardized frequency containment reserve services: the FCR services in Central Western Europe (France and Germany, Denmark-DK1 is not included in this analysis since it has joined the integrated FCR market less than a year) and Northern Europe (Denmark-DK2 and Norway).

### 4.3.1 FCR service in the Central Western Europe

When providing frequency regulation services, BESS receives remuneration mainly from the capacity payments. In exchange, BESS should provide required upward/downward energy according to the frequency deviation by discharging/charging the batteries respectively.

Table 2 presents the expected hourly required regulating energy and balancing energy for providing 1MW power capacity in FCR obtained from different sources. It should be noted that the required information for precisely simulating the energy response for frequency regulation services (e.g. detailed electricity power frequency data with time resolution in seconds) is generally not publically available.

The first three rows in Table 2 present the expected regulating and balancing energy calculated from the FCR operational data from the France market from 2019 to September 2021. The required regulating energy (FCR upward and FCR downward) is obtained directly from the published hourly operational data, and the hourly balancing energy is calculated on an hourly basis to balance the required regulating energy.

Table 2. Expected hourly required regulating energy and balancing energy for providing 1MW power capacity in the FCR Market (Germany and France)

| Country | Time Period | FCR Upward ($E_{sd}$) | FCR Downward ($E_{sc}$) | Balancing Discharging ($E_{bd}$) | Balancing Charging ($E_{bc}$) | Round-trip efficiency ($\eta$) | Notes |
|---|---|---|---|---|---|---|---|
| France | 2019 | 0.056 | 0.057 | 0.026 | 0.040 | 0.850 | Calculated from Hourly operational data from the France FCR data 2019 |
| France | 2020 | 0.051 | 0.052 | 0.022 | 0.035 | 0.850 | Calculated from Hourly operational data from the France FCR data 2020 |
| France | 2021 (Jan-Sep) | 0.054 | 0.051 | 0.020 | 0.036 | 0.850 | Calculated from Hourly operational data from the France FCR data 2021 Jan-Sep |
| Germany | 2014 | 0.035 | 0.036 | 0.019 | 0.033 | 0.776 | Thien et al.(2017) [33] |
| France | 2019 to Sep 2021 | 0.054 | 0.053 | 0.023 | 0.037 | 0.850 | Selected for evaluating the potential utilization in this study |

In Germany, the required regulating energy for the FCR service is not published by the Transmission System Operator (TSO). On the other hand, Thien et al. (2017) [33] have simulated the required energy using the power frequency data in Germany in 2014 (referenced data source for power frequency no longer available) with a 5MW BESS, presented in the fourth row in Table 2.

Comparing the expected hourly required regulating energy and balancing energy, it can be observed that the level of energy usage in France has been relatively stable during the last three years. And the regulating and balancing energy presented by Thien et al. [33] are slightly lower for Germany in 2014. The difference in energy usage between France and German data could be explained by the difference in operating strategy optimization. In the work of Thien et al. [33], the operational strategies are highly optimized for battery systems, and in contrast, the French published data is real measured data from all the service providers with different technologies. The energy usage calculated from the French data would be higher, since the efficiency for general French providers is expected to be lower than BESS with optimized operational strategy presented by Thien et al. [33].

In this study, the expected values of regulating energy and balancing energy from the French data for the last three years (presented in the sixth row in Table 2) are selected for calculating the potential utilization using the proposed model in both countries. Firstly, since France and Germany are within the same synchronous grid of Continental Europe, the frequency is assumed to be the same with usually very minor deviations. Secondly, the French data is more recent and it is stable during the years being studied. And lastly, the French data is selected to avoid overestimation of potential profitability, since the energy usage in the French data is generally higher than the data presented by Thien et al. [33], which implies a higher battery wear cost.

One may argue that in real-world operation, for any specific hour, the required regulating energy and balancing energy might be significantly different from the annual expected energy values, and this may cause overestimation/underestimation of the profitability for that specific hour. The argument is true but the proposed method is still valid since the expected energy terms are expected to be stable for a power system in the long run. And the hourly error caused by the difference between the real energy utilization and the expected energy utilization will be compensated accordingly.

Figure 5 presents the potentially profitable utilization rate for the FCR market under different battery costs for the time period 2019- Sep 2021 in France and Germany. Compared with energy arbitrage, the result shows a significant improvement in terms of potentially profitable utilization in the FCR market.

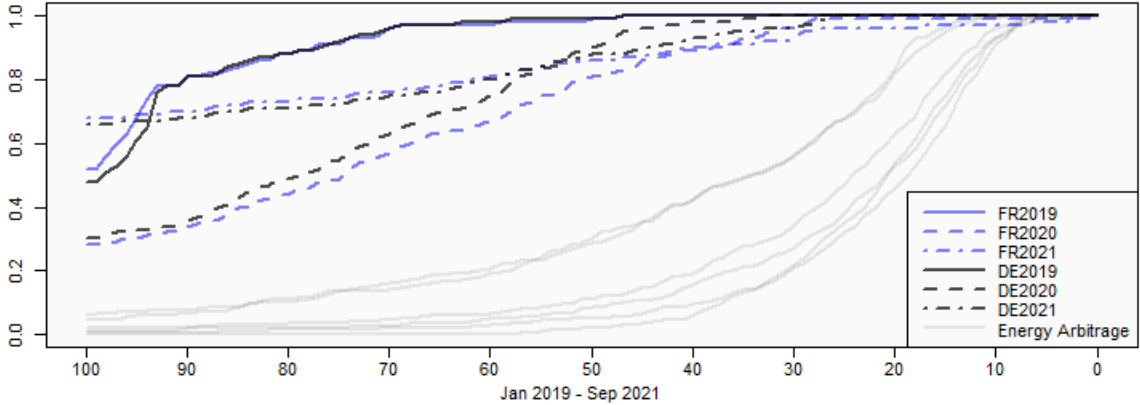

**Figure 5. Potentially profitable utilization rate (Y-Axis) for the FCR market under different battery wear cost (in €/MWh, X-axis), Jan 2019 to Sep 2021.**

**4.3.2 FCR-N service in the Northern Europe**

In Nordic countries, the FCR service is normally divided into two levels: the FCR-N (Frequency Containment Reserve for Normal operation) and FCR-D (Frequency Containment Reserves for Disturbances). In this study, the analysis is based on the FCR-N service since it provides the first-level response which requires a dynamic response to the frequency changes.

The operational information for the FCR-N service, however, is not generally publicly published for most of the markets. Starting from 2021, the Finish TSO Fingrid has published FCR-N data, including the yearly market plans, hourly market volumes, foreign trade capacity, active power reserves, and hourly market prices. Table 3 presents the expected hourly required regulating energy and balancing energy for providing 1MW power capacity in the FCR-N market in Finland from January to September 2021.

Compared with the FCR service in France and Germany, the general required energy for regulation for the FCR-N service is higher, due to the fact that the frequency band requiring full activation is lower (±0.1Hz for FCR-N in the Nordic countries and ±0.2Hz for FCR in Central Western Europe).

**Table 3. Expected hourly required regulating energy and balancing energy for providing 1MW power capacity in the FCR-N Market (Finland)**

| Country | Time Period | FCR-N Upward ($E_{sd}$) | FCR-N Downward ($E_{sc}$) | Balancing Discharging ($E_{bd}$) | Balancing Charging ($E_{bc}$) | Round-trip efficiency ($\eta$) | Notes |
|---|---|---|---|---|---|---|---|
| Finland | 2021 (Jan-Sep) | 0.091 | 0.093 | 0.079 | 0.107 | 0.850 | Calculated from Hourly operational data from the Finnish FCR-N data 2021 |

In this study, the expected values of regulating energy and balancing energy from the Finnish data in 2021 are selected for calculating the potential utilization using the proposed model for Denmark-DK2 and Norway, both of these areas, together with Finland, are within the same synchronous grid of Northern Europe.

Figure 6 presents the potentially profitable utilization rate for the FCR-N market under different battery costs for the time period 2019- Sep 2021 in Denmark and Norway. Note that the FCR-N market would include different action sessions, in this study, the daily auction D-1 is analyzed since the daily market provides more information regarding the real-time supply and demand for flexibility.

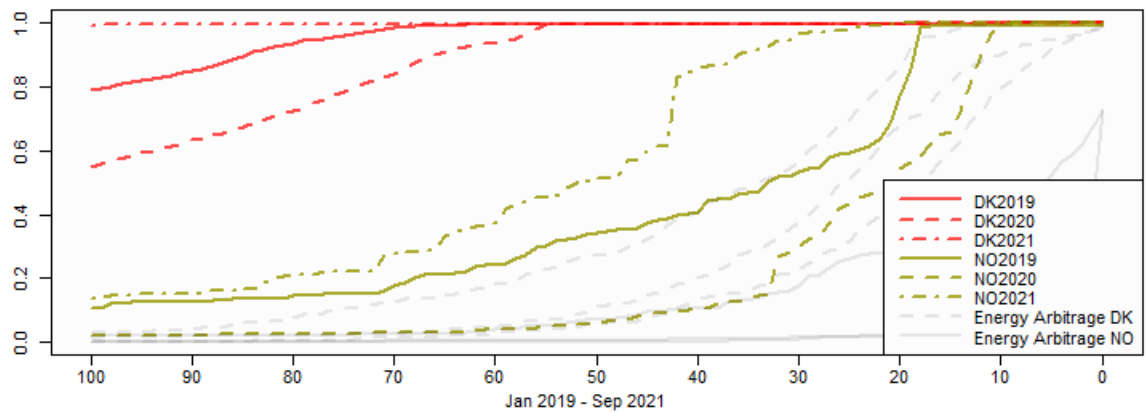

**Figure 6. Potentially profitable utilization rate (Y-axis) for the FCR-N market under different battery wear cost (in €/MWh, X-axis), Jan 2019 to Sep 2021.**

The result shows a great potential utilization for the frequency regulation service using BESS in Denmark, in January to September 2021, a BESS would be profitable in providing the FCR-N service in almost all the hourly time periods even under a marginal operating cost of 100 €/MWh. And in the other years, the potentially profitable utilization rate is also attractive under the current battery wear cost assumption.

On the other hand, the potentially profitable utilization rate for the FCR-N service in Norway is significantly lower than the Danish market. Under the current cost assumption of 100 €/MWh, the potentially profitable utilization is under 20% for all the time periods being studied. It would require further declines in the cost of the batteries for achieving a higher potential utilization. It should be also noted that, in the Norwegian market, the potential utilization of BESS in proving FCR-N service is much higher than the application of energy arbitrage.

## 5. Conclusions and discussions

Given the continuously declining cost of battery technology in the last decade, nowadays BESS becomes more attractive for power grid applications. Since BESS operation in deregulated markets would depend on the real-time market price and the operating cost, the potential utilization of BESS in different major European electricity markets has been investigated based on the proposed general BESS payoff model considering service remunerations and operational costs. Utilization factors such as potentially profitable utilization time and rate are calculated for common applications including energy arbitrage and FCR services using historical market information.

The results show that under the current empirical estimation of the installation cost and lifetime (300 €/kWh and 3000 Full equivalent cycles), the battery wear cost resulting from degradation would prevent BESS from being profitable for energy arbitrage in most of the European electricity markets. Meanwhile, significant heterogeneity of the potential profitability of BESS has been observed among different major European markets/countries. The analysis of energy arbitrage applications in the major European day-ahead markets also reveals useful information about the general scarcity of flexibility among the electricity markets. According to the result, the electricity market in Great Britain would be most promising for BESS in energy arbitrage, and similar local markets would include, but not limited to, the energy market in Ireland, and the Sicily Island of Italy. On the contrary, the potential profitability of energy arbitrage using BESS is quite limited in countries such as Norway and Spain, where the flexibility is generally provided with other energy sources such as hydropower and pumped hydro storage systems.

On the other hand, rather than buying and selling energy directly in the wholesale electricity market, BESS shows clearly and significantly higher potential in providing frequency supporting services in all the markets being studied. The result shows that even under the current battery cost assumptions, when the frequency containment reserve is remunerable, the potentially profitable utilization of BESS has become already accretive in most of the European countries, except Norway. For example in some particular time periods such as January to September 2021, the potentially profitable utilization rate may reach almost 100% for the FCR-N service in the Danish market.

In summary, comparing the major electricity markets in Europe, BESS has shown its potential in becoming a feasible solution in Central Western Europe and parts of Northern Europe by providing frequency regulation services. Further effort should be conducted focusing on the frequency regulating services in these markets to break down the barriers to BESS. On the other side, the potential application of BESS is quite limited in the Norwegian market. The most obvious explication would be the extremely high penetration level of hydropower in this country. Lastly, in other markets where the FCR services (or primary reserves) are not remunerable, such as the Apennine Peninsula and the Iberian Peninsula, BESS would not be feasible for energy arbitrage under the current situation. The potential utilization of BESS should be investigated for other frequency supporting services such as the aFRR services (secondary reserve) and mFRR services (tertiary reserve). Lastly, in the British Isles and some other islanded local markets, BESS would be considerably promising given the high volatility of the wholesale market prices, further works should be conducted to review the market design and correctly address the flexibility of BESS.

During the investigation, it has been observed that serious improvement is necessary regarding general market information accessibility and market data quality and transparency. Due to the distributive and decentralized nature of BESS, market and information transparency would be critical for the future development of BESS applications, especially for smaller market participants. For market and system regulators and operators, more efforts are necessary to ensure an open, fair, and non-discriminative market of system flexibility. Also, future research work should be elaborated by exploring novel market and service designs that adequately and appropriately address the flexibility provided by different technologies including the batteries.

This paper focuses on evaluating and comparing the potential utilization of BESS in the European electricity markets. It should be noted that historical information has been applied to study and understand the characteristics of the selected markets. Indeed, for quantitatively accessing the project value and feasibility, valuation models should be conducted for real future market scenarios. Given the fact that the proposed BESS payoff model is composed of a set of spread options based on the price difference between the service remuneration and the battery operating cost, future research of BESS feasibility may focus on valuation methods based on option pricing models.

Table 1. Summary of the FCR services in the major European electricity markets

| | Spain [50] | Italy [51] | France [52] | Germany [53] | UK [54] | Denmark [55] | | | Norway [56] | |
|---|---|---|---|---|---|---|---|---|---|---|
| Bidding Zone | ES | IT | FR | DE-LU | GB | DK1 | DK2 | | NO1-NO5 | |
| Product Name | Primary regulation | Primary regulation | FCR | FCR | Firm Frequency Response (FFR) | FCR | FCR-N | FCR-D | FCR-N | FCR-D |
| Capacity payment | None | None | Yes | Yes | Yes | Yes | Yes | Yes | Yes | Yes |
| Energy payment | None | None | As Reference spot price | None | "Response Energy Payment" Only applied when the unit are with "fuel cost" | As ordinary imbalance | Regulating power price | As ordinary imbalance | Regulating power price | None |
| Frequency response requirement | Linear response between 49.8Hz - 50.2Hz | Linear response between 49.9Hz - 50.1Hz (normal operation) | Linear response between 49.8Hz - 50.2Hz | Linear response between 49.8Hz - 50.2Hz | Linear response between 49.5Hz-50.5Hz for dynamic response Full response for frequency lower than 49.7Hz or higher than 50.3Hz for static response | Linear response between 49.8Hz - 50.2Hz | Linear response between 49.9Hz - 50.1Hz | Response between 49.9Hz (0%) to 49.5Hz (100%) for upward product Downward product be implemented in 2021 | Linear response between 49.9Hz - 50.1Hz | Response between 49.9Hz (0%) to 49.5Hz (100%) for upward product Downward product be implemented in 2021 |
| Response time | Within 15 Seconds | Within 15 Seconds | Within 15 seconds, up to 15 minutes for Limited Energy Reservoirs | Within 30 seconds, up to 30 minutes | Within 10 Seconds, up to 30 minutes for "secondary response" providers and indefinite for "high frequency response" providers | Within 15 seconds up to 30 minutes | Within 150 second up to indefinite | Supply 50% of the response within 5 seconds and the remaining within 30 seconds. | 63% of the reserve within 60 seconds and the remaining within 180 seconds, up to 15 minutes | Supply 50% of the response within 5 seconds and the remaining within 100 seconds. |
| Payment method | Non-remunerable | Non-remunerable | Marginal pricing | Marginal pricing | Pay-as-bid by monthly tenders | Marginal pricing | Pay as bid | Pay as bid | Pay as bid | Pay as bid |